\documentstyle[11pt,newpasp,twoside]{article}
\def\edcomment#1{\iffalse\marginpar{\raggedright\sl#1\/}\else\relax\fi}
\reversemarginpar

\begin{document}
\title{UV-IR Science Prospects with TES Imaging Arrays}
 \author{R.W. Romani, J. Burney, P. Brink, B. Cabrera, P. Castle}
\affil{Dept. of Physics, Stanford University, Stanford, CA 94305-4060}
\author{T. Kenny, E. Wang}
\affil{Dept. of Mech. Eng., Stanford University, Stanford, CA 94305-4021}
\author{B. Young}
\affil{Dept. of Physics, Santa Clara University, Santa Clara, CA 95053}
\author{A.J. Miller and S.W. Nam}
\affil{NIST, M.C. 814, 325 Broadway, Boulder, CO 80305}

\begin{abstract}
We are developing photon-counting cameras employing 
cryogenic arrays of energy-resolving TES (Transition Edge Sensor) pixels.
These are being tested in ground-based instruments, but will have their 
greatest impact when employed on space platforms, where they can 
cover the 10$\mu$m-100nm range with high time- and
moderate energy- resolution. Here we summarize briefly existing device 
performance, current directions in array camera development and anticipated
capabilities.
\end{abstract}

\section{Introduction and Present UV-IR TES performance}

	Cryogenic energy-resolving photon detectors show considerable
promise for advanced instrumentation in several wavebands. In 
the near-IR through UV, these devices provide noise-free photon 
counting with $\le \mu$s time resolution at $\delta E \approx 0.15$eV
energy resolution. Early work focused on superconducting tunnel junction
devices (Perryman, Foden \& Peacock 1993; Peacock, these proc.),
but alternative cryogenic technologies show great
promise. Our group has pursued the application of superconducting TES
devices to this problem (Cabrera et al. 1998). Our detectors are pixel arrays of
40nm thick tungsten (W) films patterned on Si. When cooled below their 
$\sim 100$mK
transition temperature and voltage biased to produce negative electro-thermal 
feedback (Irwin 1995), an absorbed photon decreases the Joule-heating current.
This pulse signal is read out with a DC
SQUID array (Welty \& Martinis 1993). Bi-layer thin films, a variety of
substrates, and multiplex schemes to read out larger arrays are also
under development and TES array technology is finding interesting application
from the sub-mm (SCUBA-2) to the X-ray (Con-X) regimes.

Miller et al. (2000) summarize performance of present W TES devices. 
We routinely achieve $\delta E\sim 0.15$eV
at $\sim 3$eV, absolute GPS photon times to 300ns and 
single-pixel count rates of $\sim 30$kHz. Energy discrimination rejects
all cosmic rays and allows no `dark current' above the noise floor.
The intrinsic bare W quantum efficiency is $\sim 50$\% and in astronomical
systems we have achieved efficiencies on the sky as high as $\sim$20\%.

TES resolution is given by
$\Delta E_{FWHM} = 2.355 [ 4k_BT^2C(n/2)^{1/2}/\alpha ]^{1/2}$,
where $T$ and $C$ are the temperature and heat capacity of the W $e^-$ system,
$n=5$ for electron-phonon conduction and $\alpha = (d {\rm ln}R/d{\rm ln}T)_
{V=const}$. The pixel size sets $C$; the maximum
(saturation) energy is $E_{max} \approx CT_c/\alpha$ and 
therefore the optimum
detector design gives $\Delta E_{FWHM,min}=2.355[4k_BT(n/2)^{1/2}E_{max}
]^{1/2}$.  Thus for our typical 20$\mu$m pixel and $E_{max} = 10$eV we 
obtain $\Delta E_{FWHM,min} \approx 0.05$eV. Somewhat higher resolution 
can be obtained with lower $E_{max}$ by operating in the saturation regime, 
measuring pulse shape rather than peak height.  In the 
present devices on Si, 58\% of the photon energy is lost from the W $e^-$
system to substrate phonons, giving an expected $\Delta E =$0.088eV; 
for comparison the measured resolution is
0.12eV FHWM.  Interestingly, although the physics is different, these 
sums predict $\Delta E_{FWHM,min}$ quite similar to theoretical limits 
and goal resolutions for existing STJ devices.  The eventual choice of detector 
is probably best decided on the basis of achieved performance and ease of 
manufacture.  At low $E$ the thermodynamic fluctuation noise floor allows
detection to $\sim 10 \mu$m.  Of course, this full sensitivity range cannot be 
exploited from the ground, and we have employed a variety of filtering 
schemes to suppress unacceptably high thermal count rates beyond $2\mu$m.

	To illustrate the astronomical utility of these detectors, we
have packaged TES arrays into simple demonstration systems, based on a
He dilution refrigerator used for device development, and performed
some basic observations. To filter the IR background, we employ a $\sim 3$m
length of high-OH optical fiber spooled at 4K in the dewar. The OH
bands of this cold filter provided effective blocking at $\sim 1.7\mu$m,
but allow transmission in the atmospheric windows. At the telescope, the
atmosphere and optics limit us to $\sim 3.5$eV, but 
in the lab we detected $E>7$eV photons. With such systems
we demonstrated the first astronomical spectra taken with a cryogenic
optical detector 
and using small telescopes have
made unique, scientifically valuable, measurements of the Crab pulsar
and other astronomical objects (Romani et al. 1999, Romani et al. 2001), 
including photon counting spectro-photopolarimetry in the optical/IR.

	On small telescopes these test set-ups have achieved good
efficiency, although system efficiencies of $\le 10$\% are typical. 
As we are fiber-coupled we have to date only employed a small
subset of the TES array (e.g. 2$\times$3 out of $6\times6$ for the 
McDonald 2.7m Crab observations). Also, for ease of fabrication,
the present TES arrays only employ a single lithographic layer of wiring;
these wiring leads are responsive to photons, which complicates the energy PSF and
spectral calibration. Finally, fiber coupled schemes, while attractive
for test purposes, do not match well to the eventual (space platform)
imaging applications for which these TES devices should have
their largest impact.

\section{Current Efforts}

	All of these factors can be mitigated with some simple design changes,
and so we are assembling a first dedicated TES system for ground-based
optical/UV astronomy. We are currently assembling an array in
a compact, portable Adiabatic 
Demagnetization Refrigerator (ADR) with dewar windows allowing direct
imaging on the focal plane array. This solution is suitable for Cassegrain or
Nasmyth applications at moderate to large aperture ground-based telescopes.
The system accepts a slow f/10-f/20 beam and focal reduces with a cryogenic
all-reflective system to allow good coupling to the devices.
At the focal plane we read out a 32 pixel
array. The existing design may be run at up to $8 \times 8$ format by
adding more read channels; with modest changes to the device fabrication,
arrays of 4-6$\times$ this area could be made. With a window solution,
we employ heat-absorbing filters passing primarily in the
optical/UV, so the system efficiency can be increased by appropriate AR
coatings. Finally, to increase the fill factor and suppress energy PSF 
features from the wires and substrate, we will employ a focusing mask that
collects the light to the $20\times 20\mu$ W TES pixels. 

\section{Future Potential from Space Platforms}

	As ground-based tests demonstrate, these cryogenic technologies 
most immediately impact faint source observations when 
broad features in the SED and rapid time variability encode the key 
physics. Compact object studies, especially spin-powered pulsars and 
accretion-powered CVs and black hole systems are thus the `killer-app'
for first generation TES systems. Some of this science can be done from
the ground, but the broad (0.1-10eV) energy window accessed from space 
presents opportunities for uniquely powerful investigations.

	The good QE and noise properties of these devices encourages
application to a wider range of problems, especially in the UV. Clearly,
the main driver here is array format. Imaging applications 
providing a low-resolution ultra-broad band spectrum of extragalactic
sources (see Romani et al. 1999, Mazin \& Brunner 2000) over a modest field 
are very appealing.
However, these arrays may also be useful in different formats such as
order-sorting detectors for echelle spectroscopy in
the optical/UV. The ability to read out more than the modest $\sim 10^2$
simultaneous channels covered in existing cryogenic system designs is
therefore quite important. 
One fundamental limit is the complexity and heat load of the wiring 
delivering the signals to room temperature.
In the case of TES devices there is a demonstrated
(Chervenak et al 1999) scheme for multiplex readout at moderate count rates
of a number of TES pixels by a single SQUID channel. Multiplex
technology development is also being driven by other TES applications and
should certainly increase the useful arrays sizes by factors of 
$~\sim 30 -10^3$. There are other schemes for allowing large cryogenic 
pixel count without a prohibitive number of wires to low temperature
(e.g. `kinetic inductance' coupling; Zmuidzinas et al. 2002).

	What is needed to give TES imaging arrays (or
comparable cryogenic technology) an exciting science role on a 
space-based platform? We envision three levels of 
capability. The first stage is a small $\sim 8\times8 - 16\times 16$ 
pixel imaging array with
low ($\sim 0.1$eV) energy resolution detecting the near IR-UV band with good
QE. This already provides dramatic advance on time resolved studies
of Galactic compact objects, with a roughly PSF-matched format and
the option for local background subtraction, polarization and similar modes.
This array could also perform areal spectroscopy of small extragalactic
objects reaching remarkably faint magnitudes. It may be built as a fairly
straightforward rescaling of present systems using today's techniques. Such 
a (special purpose) system would enable very exciting science from a small 
aperture dedicated facility or a sub-orbital platform and in our opinion
is should be pursued both for the science and for the demonstration of
these detectors in a flight program.

	A more capable facility would have 1-10 kpix format and would 
require some form of multiplex read-out. These read channels could be
used both as an imaging array and in an elongated (order sorting) 
spectrographic format. It should be possible to have the same (expensive)
read channels address several (cheap) array formats in a single dewar. 
This array appears to be a natural, near term ($\sim5$y) extension of
present programs. {\it If} such an array can be coupled with a reliable
closed-cycle, space-qualified cryocooler+ADR system, a strong science case 
can be made for a dedicated small aperture system. Grid filtering and 
passive optics cooling in space could allow simultaneous access to 
$\sim$0.2-10eV and imaging and spectroscopic modes, as above. 

	The ultimate instrument would have $\ge 10^5$ pixels, which
with the high QE, ultra-broad sensitivity range, noiseless detection
and intrinsic energy resolution would exceed the capabilities 
of CCD/MAMA arrays for a variety of mainstream astrophysics and cosmology 
applications. Such an array might be dynamically addressed, reading out only
an interesting pixel subset. Light travel time
ensures that sub-$\mu$s timestamps are unlikely to be of interest
over large areas, so cryogenic preprocessing to accumulate simple
spectral images would also be desirable.  Schemes to build such
an array and to accomplish the required read-out and processing are 
presently at the cartoon stage; hence prognostication of a development
time is rather pointless. However, intermediate steps are quite compelling
so with sufficient resources, progress toward such ultimate astronomical
detectors is likely to occur.

This work was supported in part by NASA grant NAG5-3775 and made use of the 
Stanford Nanofabrication Users Network funded by NSF award ECS-9731294.

\end{document}